\documentclass[aps,superscriptaddress]{revtex4-1}

\usepackage[dvips]{graphicx}
\usepackage{amssymb,amsfonts,amsmath,color}

\newcommand{\bs}{{\bf {s}}}
\newcommand{\br}{{\bf {r}}}
\newcommand{\bv}{{\bf {v}}}
\newcommand{\bk}{{\bf {k}}}
\newcommand{\etal}{{\it et al.}~}
\newcommand{\bom}{{\mbox{\boldmath $\omega$}}}

\begin{document}

\title{Thermally- and mechanically-driven quantum turbulence in helium II}

\author{A.~W.~Baggaley}
\author{L.~K.~Sherwin}
\author{C.~F.~Barenghi}\affiliation{Joint Quantum Centre Durham-Newcastle, School of Mathematics and Statistics, University of
Newcastle, Newcastle upon Tyne, NE1 7RU, UK}

\author{Y.~A.~Sergeev}\affiliation{Joint Quantum Centre Durham-Newcastle, School of Mechanical and Systems Engineering, Newcastle
University, Newcastle upon Tyne, NE1 7RU, UK}

\begin{abstract}
In most experiments with superfluid helium turbulence is
generated thermally (by applying a heat flux, as in thermal counterflow) 
or mechanically (by stirring the liquid).  By modelling the superfluid
vortex lines as reconnecting space curves with fixed circulation, 
and the driving normal fluid as a uniform flow (for thermal
counterflow) and a synthetic turbulent flow (for mechanically-driven
turbulence), we determine the difference between thermally-driven and 
mechanically-driven quantum turbulence.
We find that in mechanically-driven turbulence
the energy is concentrated at the large scales,  the spectrum obeys
Kolmogorov scaling, vortex lines have large curvature, and the presence
of coherent vortex structures induces vortex reconnections
at small angles. On the contrary, in thermally-driven turbulence the energy is 
concentrated at the mesoscales, the curvature is smaller, the
vorticity field is featureless and reconnections occur at larger angles.
Our results suggest a method to experimentally
detect the presence of superfluid vortex bundles.
\end{abstract}


\maketitle

\section{Introduction}
\label{section:1}

Recent work \cite{Skrbek-Sreeni,Vinen-Niemela} 
has highlighted similarities between the turbulence of superfluid 
helium~II (quantum turbulence) and the turbulence of 
ordinary (classical) fluids.  In particular,
experimental \cite{Tabeling,Salort} and theoretical
\cite{Nore,Araki,Kobayashi,Sasa-Tsubota,Baggaley-fluctuations,
Baggaley-structures,Lvov} studies have established that
the distribution of the superfluid kinetic energy over the length scales
(energy spectrum)
obeys the same $k^{-5/3}$ Kolmogorov scaling of ordinary turbulence
\cite{Frisch} 
where $k$ is the wavenumber. The similarity is remarkable 
because helium~II is unlike an ordinary fluid:
firstly, it has a two-fluid nature, consisting of 
a viscous normal fluid component and an inviscid
superfluid component coupled by a mutual friction \cite{Donnelly};
secondly, superfluid vorticity is not a continuous field (like in
an ordinary fluid) but is restricted to discrete vortex filaments 
around which the circulation is fixed to the
ratio of Planck's constant and the mass of one helium atom.

In most experiments turbulence in helium~II is excited mechanically 
(by stirring the helium with grids or propellers 
\cite{Oregon,Tabeling,Salort} or forcing it along pipes
\cite{VanSciver}),  or thermally (by the application of
a heat flux \cite{Vinen,Tough}). The simplest, most studied form of
thermal stirring (to which hereafter we restrict our work) is called
thermal counterflow. The name arises because the
normal fluid and the superfluid move in opposite directions,
so that their velocity difference is proportional
to the applied heat flux and the net mass flux is zero.
Other forms of heat transfer (e.g. pure superflow)
and other techniques to generate
turbulence (e.g. ultrasound \cite{CWSmith}, ion injection and 
spin-downs \cite{Walmsley})
are either less studied, or refer to the low temperature limit
(below $1~\rm K$), or have a special character 
(rotating turbulence, turbulent fronts, the Kibble-Zurek mechanism, etc.).
thus they are not our interest here, and nor are the special methods
used to model them;
for these aspects we refer the reader to a recent 
review \cite{Halperin-Tsubota}.

The aim of this work is to clarify the difference
between thermally-excited counterflow turbulence
and mechanically-excited turbulence in helium~II.  For simplicity, we are
concerned only with statistical steady state turbulence away from boundaries
(thus ignoring the important problems of turbulence decay
and flow profiles), and at the relatively high temperatures,
where mutual friction plays a role. 

After setting up the necessary numerical models (section~\ref{section:2}) 
we compute for the first time
the energy spectrum of counterflow
turbulence, compare it to the spectrum of mechanically-driven turbulence,
and test the idea which has been proposed
in the literature that ``counterflow turbulence has only
one length scale'', meaning the average intervortex distance 
(section~\ref{section:3}).
We also find that thermally- and mechanically-induced turbulence
differ with respect to curvature (section~\ref{section:4}), the presence of
coherent structures (section~\ref{section:5}) 
and vortex reconnection statistics.
The last result suggests a method to detect experimentally
the existence of superfluid vortex bundles (section~\ref{section:6}). 
Section~\ref{section:7} summarises the conclusions.

\section{Numerical Method}
\label{section:2}

Following Schwarz \cite{Schwarz}, we model quantum vortex filaments
as space curves $\bs(\xi,t)$ which move according to

\begin{equation}
\frac{d{\bf s}}{dt}=\bv_s^{tot}+\alpha \bs' \times (\bv_n^{ext}-\bv_s^{tot})
-\alpha' \bs' \times (\bs' \times (\bv_n^{ext}-\bv_s^{tot})),
\label{eq:Schwarz}
\end{equation}

\noindent
where $t$ is time, $\alpha$ and $\alpha'$ are temperature dependent friction
coefficients \cite{Donnelly-Barenghi}, $\bs'=d\bs/d\xi$ is the unit
tangent vector at the point $\bs$, $\xi$ is arc length, and
$\bv_n^{ext}$ and $\bv_s^{ext}$ are externally applied normal fluid and
superfluid velocities.  The self-induced velocity
of the vortex filament at the point $\bs$ is given by
the Biot-Savart law \cite{Saffman}
\begin{equation}
\bv_s^i=
-\frac{\kappa}{4 \pi} \oint_{\cal L} \frac{(\bs-\br) }
{\vert \bs - \br \vert^3}
\times {\bf d}\br,
\label{eq:BS}
\end{equation}

\noindent
where $\kappa=9.97 \times 10^{-4}~\rm cm^2/s$
is the quantum of circulation and the line integral
extends over the entire vortex configuration $\cal L$. 
The total superfluid velocity is thus $\bv_s^{tot}=\bv_s^i+\bv_s^{ext}$.

In the case of thermal counterflow (which hereafter we simply refer to as
thermally-driven, to contrast it to mechanically-driven turbulence),
$\bv_n^{ext}$ and $\bv_s^{ext}$
arise from the imposed heat flux $\dot Q$, where
$v_{ns}=\vert \bv_n^{ext} - \bv_s^{ext} \vert={\dot Q}/(\rho_s S T)$ is the
counterflow velocity, $v_n^{ext}={\dot Q}/(\rho S T)$,
$v_s^{ext}=-\rho_n v_n^{ext}/\rho_s$, $S$ is the specific entropy, $T$ is
the temperature, $\rho_s$ and $\rho_n$ are the superfluid and
normal fluid densities, and $\rho=\rho_s+ \rho_n$. 
We make the usual \cite{Schwarz,Adachi} simplifying
assumption of uniform velocity profiles $\bv_n^{ext}$ and $\bv_s^{ext}$
away from boundaries, neglecting the possibility
that at sufficiently large ${\dot Q}$ the normal fluid becomes
turbulent \cite{Melotte,McKinsey}. 
It is also convenient to perform the calculation in the frame
of reference of the imposed superflow, setting $\bv_s^{ext}=0$.

In the case of mechanically-induced turbulence, since we ignore
boundaries and flow profiles, we set
$\bv_s^{ext}={\bf 0}$ in Eq.~(\ref{eq:Schwarz}) and
replace $\bv_n^{ext}$ with the following 
synthetic turbulent flow \cite{Osborne}:

\begin{equation}
\bv_n^{ext}(\bs,\,t)=\sum_{m=1}^{m=M}({\bf A}_m \times \bk_m \cos{\phi_m}
+{\bf B}_m \times \bk_m \sin{\phi_m}),
\label{eq:KS}
\end{equation}

\noindent
where $\phi_m=\bk_m \cdot \bs + \omega_m t$, $\bk_m$ and 
$\omega_m=\sqrt{k^3_m E(k_m)}$ are  wave vectors and angular frequencies.
This $\bv_n^{ext}$ is solenoidal, time-dependent, and,
with a suitable choice of ${\bf A}_m$ and ${\bf B}_m$ 
(adapted to the periodic box \cite{Wilkin}), its
energy spectrum has Kolmogorov form $E(k_m)\sim k_m^{-5/3}$
in the range from $k_1$ (corresponding to the integral scale) to $k_M$
(corresponding to the dissipation scale). Synthetic turbulence is widely
used to study transport properties, and compares very well with
direct numerical simulations and experiments (for example it satisfies
observed two-points turbulence statistics). 

Our calculations are performed in a periodic cube of size $D=0.1~\rm cm$.
The numerical techniques to discretize the vortex filaments
into a variable number of points $\bs_j$ ($j=1, \cdots N$) held
at minimum separation $\Delta\xi/2$, compute the time
evolution, de-singularize the Biot-Savart integrals, evaluate $\bv_s^i$
using a tree-method (with critical opening angle $0.4$), and 
algorithmically perform vortex reconnections when vortex lines
come sufficiently close to each other, are all
described in our previous 
papers~\cite{Baggaley-cascade,Baggaley-fluctuations,Baggaley-tree,
Baggaley-reconnections}.

It must be stressed that our models have a
limitation: the normal fluid is prescribed
rather than computed self-consistently.
The inclusion of the back-reaction of the superfluid vortices onto
the normal fluid would require the numerical solution of the Navier-Stokes
equation for the normal fluid (suitably modified by the inclusion of a
mutual friction term), alongside the time evolution of the superfluid
vortices. However, a dynamically self-consistent model would be
very complex and computationally expensive, and one
could not easily explore parameter space and the effects of changing
numerical resolution and initial conditions. This approach was attempted 
for a single vortex ring \cite{Kivotides-science}. In the case of
turbulence, this approach has so far been limited to the initial growth of a cloud
of vortex lines \cite{Kivotides-coupled}.

\section{Energy spectrum}
\label{section:3}

\begin{figure}[h]
\begin{center}
\includegraphics[width=0.34\textwidth]{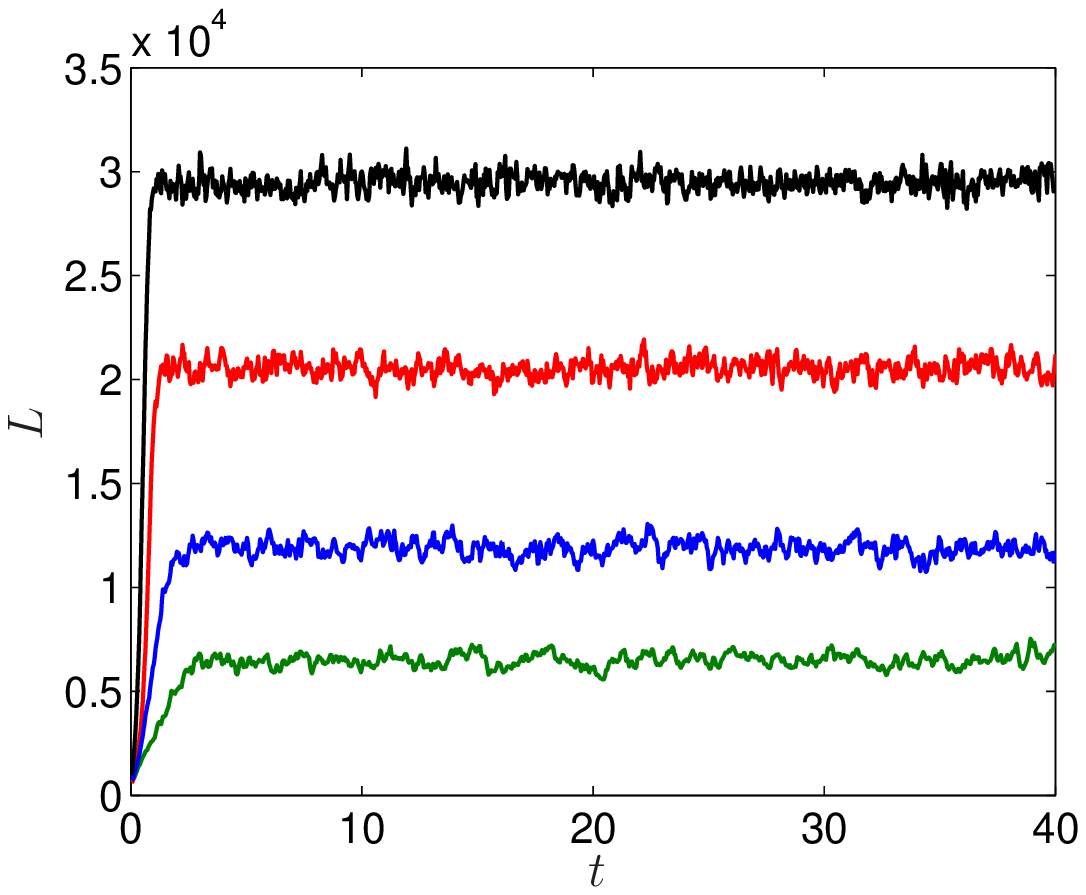}\\
\caption{
(Color online) Thermally-induced turbulence.
The evolution of the vortex line density $L$ (cm$^{-2}$) vs time $t$ (s)
at counterflow velocities (from top to bottom)
$v_{ns}=1.25~\rm cm/s$ (black), $1.0~\rm cm/s$ (red), 
$0.75~\rm cm/s$ (blue), and $0.55~\rm cm/s$ (green).
}
\label{fig:1}
\end{center}
\end{figure}

\begin{figure}[h]
\begin{center}
\includegraphics[width=0.34\textwidth]{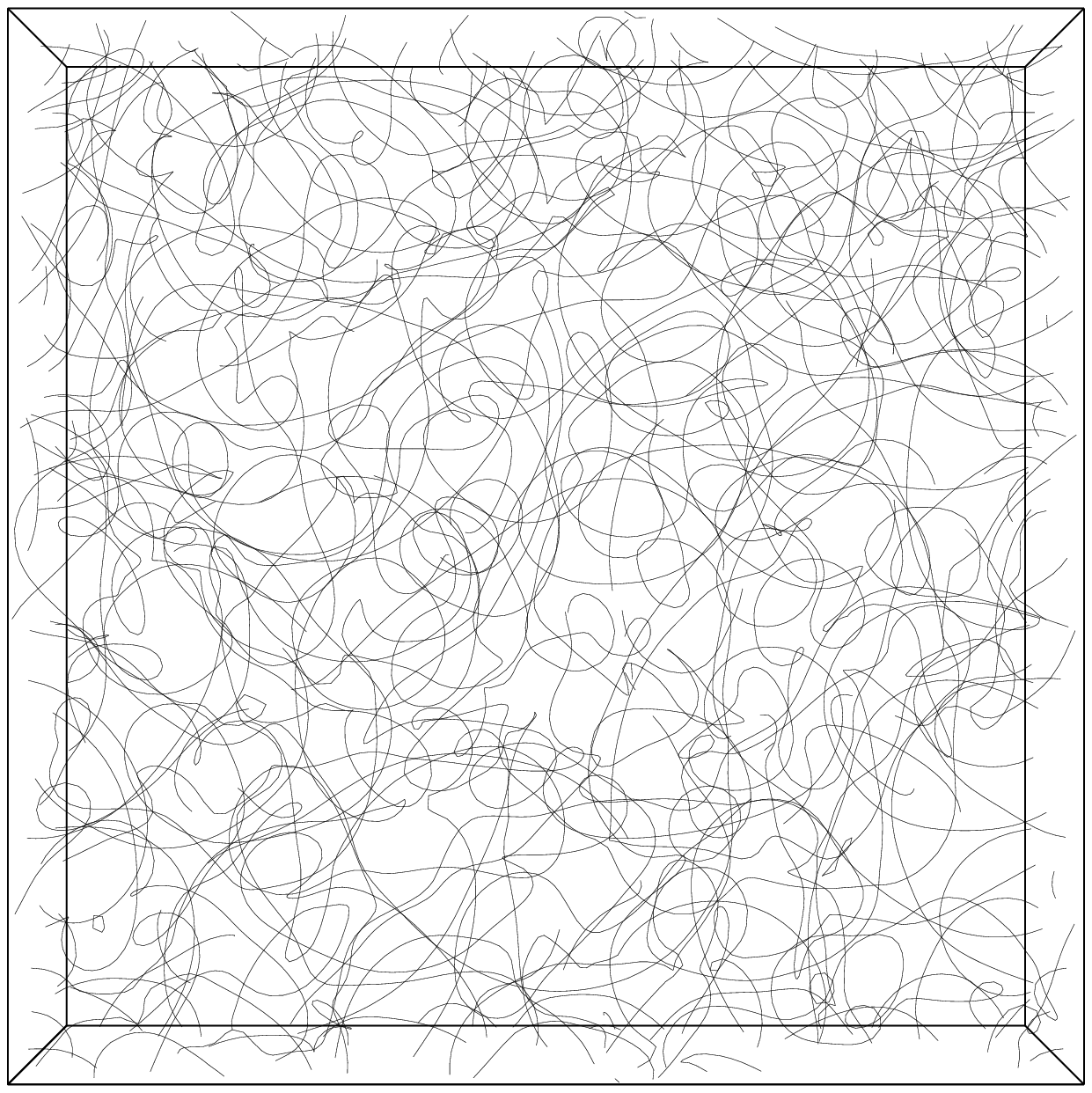}\\
\includegraphics[width=0.34\textwidth]{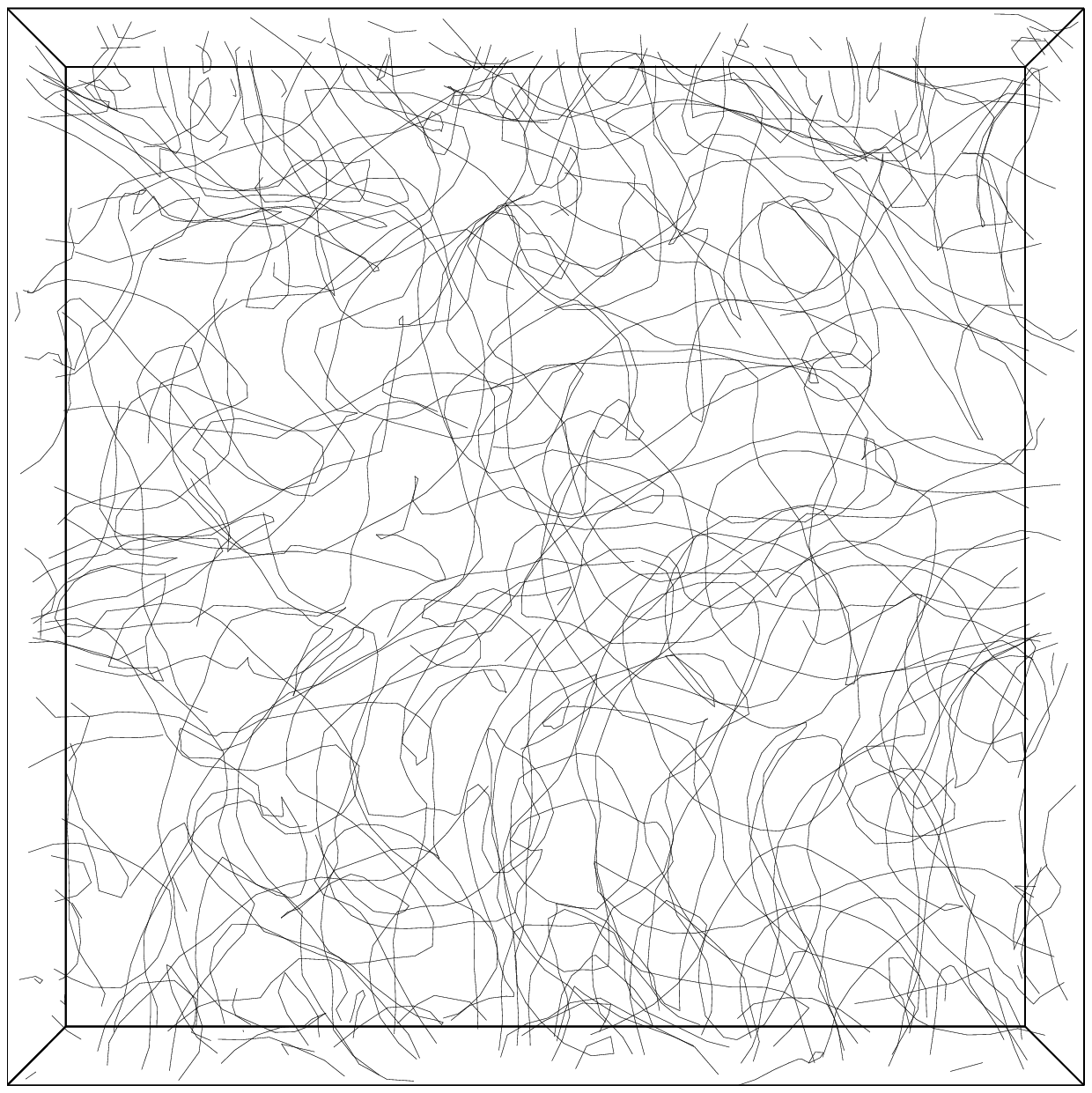}
\caption{
Snapshots of vortex tangles ($y,z$ projections). 
Top: thermally-driven by counterflow 
($v_{ns}=0.75~\rm cm/s$, $L \approx 12000~\rm cm^{-2}$);
bottom: mechanically-driven
($Re=208$, $L \approx 12000~\rm cm^{-2}$)
}
\label{fig:2}
\end{center}
\end{figure}

We choose temperature $T=1.9~\rm K$
(at which $\alpha=0.206$ and $\alpha'=0.0083$) 
which is typical of 
experiments and allows direct comparison with
previous work, and numerical resolution $\Delta\xi=0.0016~\rm cm$.
First we calculate thermally-induced turbulent vortex tangles at increasing
values of $v_{ns}$. We find that, after an
initial transient, the vortex line density $L$ (vortex length per
unit volume) saturates to a statistically steady 
state (see Fig.~\ref{fig:1}) of density $L=\gamma^2 v_{ns}^2$
which is independent of the details of the
initial condition (various vortex loops configurations were tried).
Fig.~\ref{fig:2} (top) shows a snapshot of such vortex tangle.

Our vortex line densities agree 
with previous work;
for example, taking $T=1.9~\rm K$ we obtain 
$\gamma \approx 137~\rm s/cm^2$
which compares well to $ \gamma \approx 140$ and $133$ obtained 
in the numerical simulations of Adachi \emph{et al.}~\cite{Adachi} and in the
experiments of 
Childers and Tough~\cite{Childers} respectively.

\begin{figure}[h]
\begin{center}
\begin{tabular}{cc}
\includegraphics[width=0.50\textwidth]{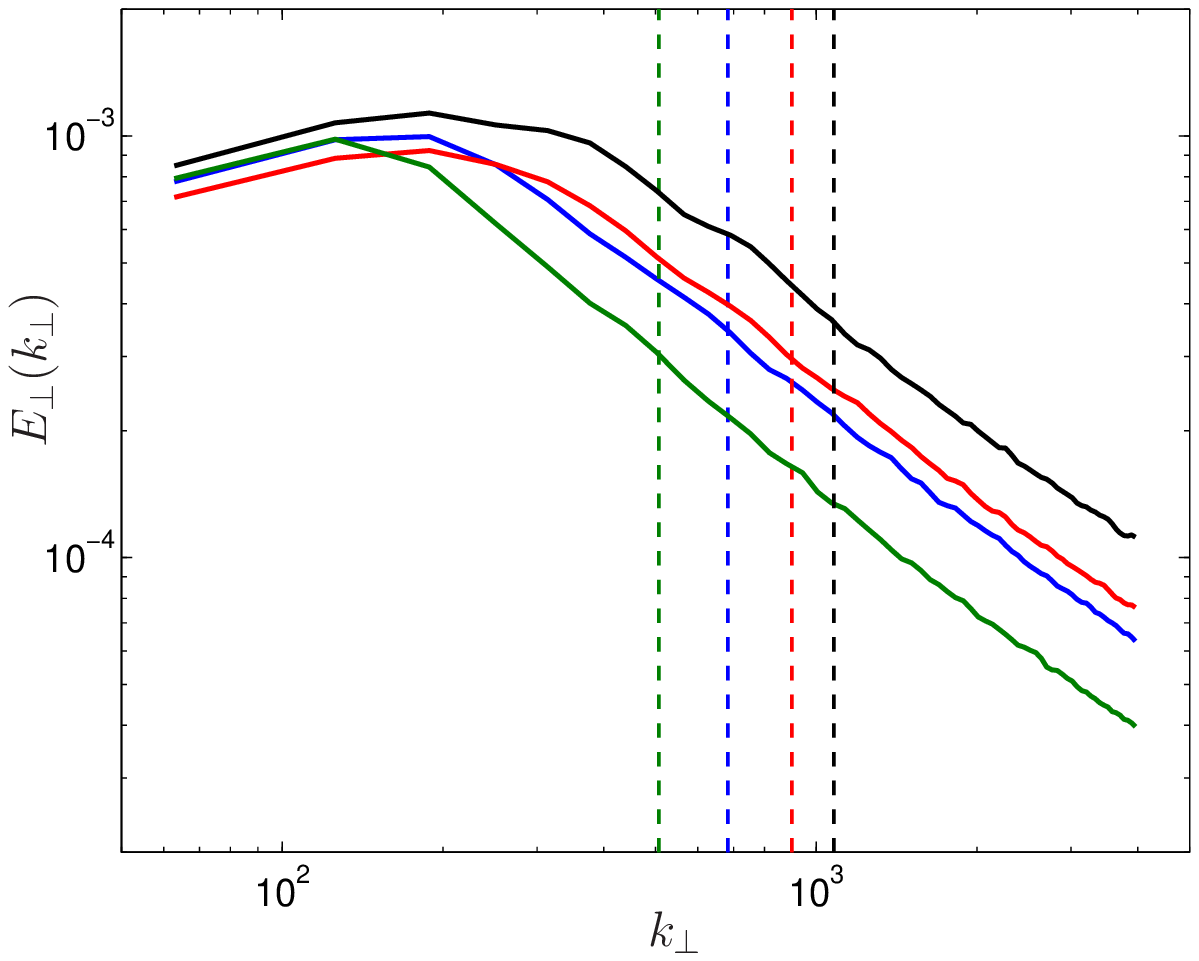}\\
\includegraphics[width=0.50\textwidth]{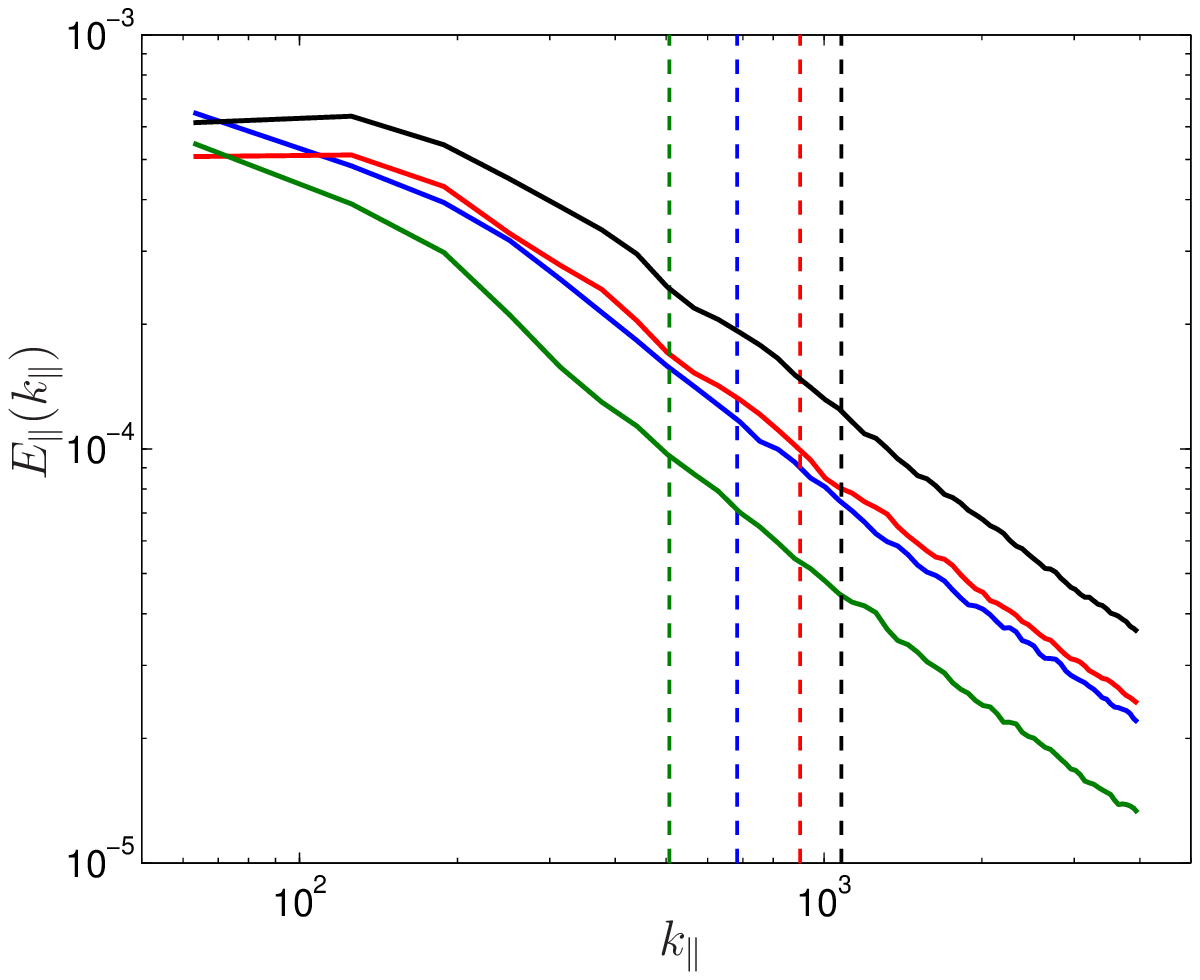}
\end{tabular}
\caption{
(Color online)
Counterflow turbulence.
Top: the perpendicular energy spectrum $E_\perp(k_\perp)$ (arbitrary units) vs wavenumber $k_\perp$ (${\rm cm^{-1}}$).
Bottom: the parallel energy spectrum $E_{\parallel}(k_\parallel)$ (arbitrary units) vs wavenumber $k_\parallel$ (${\rm cm^{-1}}$).
The vertical lines mark $k_{\ell}$ at increasing
$v_{ns}$ from right to left.
}
\label{fig:3}
\end{center}
\end{figure}

To analyze our results we Fourier-transform the superfluid velocity and
compute the energy spectrum $E_k$. If the turbulence is 
isotropic, $E_k$ is defined by

\begin{equation}
E=\frac{1}{V} \int_V \frac{1}{2} \bv_s^2 dV=\int_0^{\infty} E_k dk,
\label{eq:Ek}
\end{equation}

\noindent
where $V$ is volume,
$k=\vert \bk \vert$, and $\bk$ is the three-dimensional wavenumber.
However, it is well known \cite{Schwarz,Adachi} that counterflow 
turbulence is flattened on the ($y,\,z$) plane perpendicular 
to the direction ($x$) of the heat flux. For example, if
$L_x$, $L_y$ and $L_z$ are the vortex lengths (per unit
volume) projected in the $x$, $y$ and $z$-direction,
at $v_{ns}=1.25~\rm cm/s$ we have $L_x/L=0.34<L_y/L=0.55=L_z/L$.
It is therefore better to distinguish between parallel and
perpendicular superfluid energy spectra,
$E_{\parallel}$ and $E_{\perp}$, calculated replacing 
$\bv_s^2=v_{sx}^2+v_{sy}^2+v_{sz}^2$ in Eq.~(\ref{eq:Ek}) 
with $3 v_{sx}^2$ and
$(3/2)(v_{sy}^2+v_{sz}^2)$ respectively. 
Fig.~\ref{fig:3} (top) shows $E_{\perp}$  
for various $v_{ns}$ plotted in the range 
$k_D=2 \pi /D \le k_{\perp} \le k_{\Delta\xi}=2 \pi/\Delta\xi$
(where $k_{\perp}$ is the perpendicular wavevector);
the vertical lines mark the wavenumbers $k_{\ell} =2 \pi/\ell$ 
corresponding to the average intervortex spacing, $\ell \approx L^{-1/2}$.
It is apparent that the perpendicular energy spectrum $E_{\perp}$ has
a broad peak in the mesoscales at intermediate wavenumbers
$k_D< k<k_{\ell}$. At larger $k$ the spectrum follows
the typical $k^{-1}$ scaling of smooth isolated vortex lines as
expected. The parallel spectrum $E_{\parallel}$ vs $k_{\parallel}$
(where $k_{\parallel}$ is the parallel wavevector)
exhibits similar 
features, see Fig.~\ref{fig:3} (bottom). 
Plotting $E_{\perp}$ and $E_{\parallel}$ vs $k$ rather than $k_{\perp}$
and $k_{\parallel}$ yields similar results. We note that the
 counterflow energy spectrum, which we measure, is qualitatevely similar
to the spectrum shown by Nemirovskii, Tsubota and Araki \cite{Nemirovskii-Tsubota-Araki-2002} in their Fig.2.

\begin{figure}[h]
\begin{center}
\includegraphics[width=0.50\textwidth]{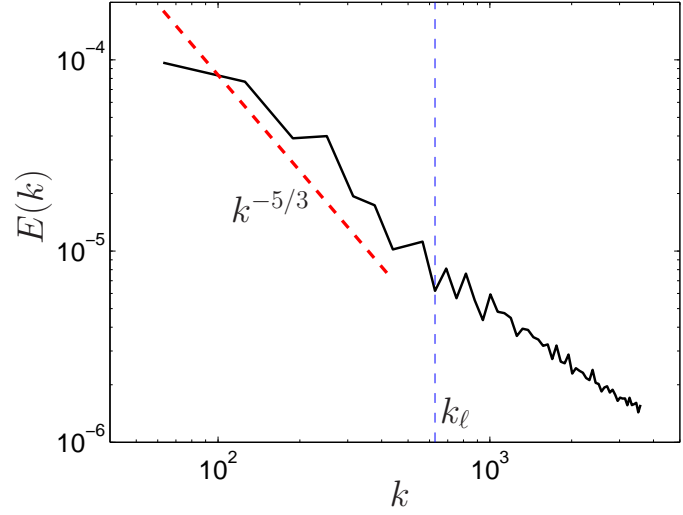}
\caption{
(Color online)
Mechanically-induced turbulence.
Energy spectrum $E_k$ (arbitrary units) vs wavenumber $k$ (${\rm cm^{-1}}$)
of vortex tangle driven by the synthetic turbulent flow of
Eq.~(\ref{eq:KS}) with $M=188$ modes. The vertical dashed blue 
line marks $k_{\ell}$. The dashed red line shows the $k^{-5/3}$ Kolmogorov
scaling.
The effective Reynolds number of the normal fluid is
$Re=(k_M/k_1)^{4/3}=208$, defined by the condition that the
dissipation time equals the eddy turnover time at $k_M$.
}
\label{fig:4}
\end{center}
\end{figure}

Proceeding in analogy to what we did for counterflow turbulence, 
we start from an arbitrary seeding initial condition, drive the vortex
tangle with the synthetic turbulent flow of Eq.~(\ref{eq:KS}),
and let $L$ grow and saturate to a statistical
steady state of turbulence which does not depend on the initial
condition (the time behaviour of $L$ is similar to Fig.~\ref{fig:1}).
A snapshot of this mechanically-driven tangle
is shown in Fig.~\ref{fig:2} (bottom). We then
compute the superfluid energy spectrum. In agreement with previous
experimental \cite{Tabeling,Salort} and theoretical
\cite{Baggaley-fluctuations,Lvov} work, we find
that the energy is concentrated at
the largest scales, $k \approx k_D$,
 and that $E_k \sim k^{-5/3}$ for large $k$
(see Fig.~\ref{fig:4}).

We conclude that there is a remarkable spectral difference between 
thermally-driven turbulence and mechanically-driven turbulence.
Whereas in the former the turbulent kinetic energy is
concentrated at intermediate length scales, in the latter most of
the energy is at the largest scales, as in classical ordinary turbulence. 

An argument is often made in the literature that counterflow
turbulence has only one characteristic length scale, the intervortex
distance $\ell$: it is apparent from Fig.~\ref{fig:3} that $E_k$ does not
have a sharp peak at $k \approx k_{\ell}= 2 \pi /\ell$ 
(indicated by the vertical lines), 
but rather a broad maximum at smaller wavenumbers in the mesoscale region
$k_D < k \le k_{\ell}$. The traditional argument, although quantitatively
wrong, is thus qualitatively correct.

\section{Curvature}
\label{section:4}

\begin{figure}[h]
\begin{center}
\includegraphics[width=0.50\textwidth]{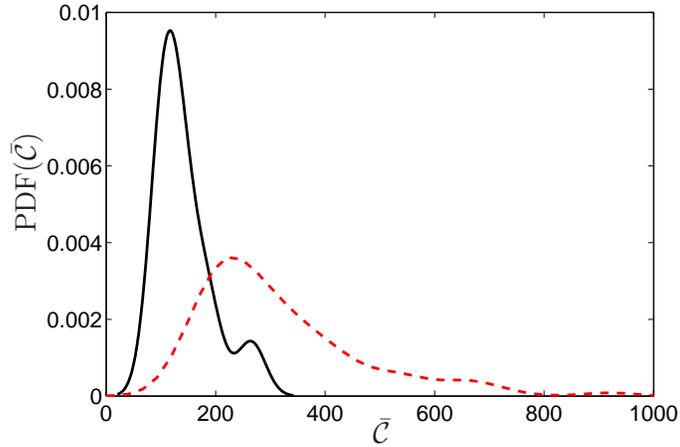}
\caption{
Probability density function (PDF) of the mean curvature
per vortex loop $\bar{C}$. Solid black line: mechanically-driven turbulence; 
dashed red line: thermally-driven turbulence. Notice the larger curvatures
present in thermally-driven turbulence.
}
\label{fig:5}
\end{center}
\end{figure}

If we look carefully at the vortex tangles shown in Fig.~\ref{fig:2},
we notice that the thermally-driven tangle (top) contains relatively
more closed loops, and the mechanically-driven tangle (bottom) contains
relatively more long vortices which extend throughout the
periodic  computational
domain.
We sample the curvature $C=\vert \bs'' \vert$ along
each vortex loop
and construct the probability density function (PDF) of the mean  
curvature $\bar{C}$ of each distinct loop. Fig.~\ref{fig:5} shows
the result.
We notice that mechanically-driven turbulence 
contains smaller curvatures (that is, larger radii of curvature $R=1/C$)
than thermally-driven turbulence; indeed, for the latter
${\rm PDF}({\bar C})$ has a maximum at ${\bar C} \approx 250$
in correspondence of the maximum of the energy spectrum 
shown in Fig.~\ref{fig:3}.

As an additional numerical experiment,
we compute the energy spectra of 
configurations of circular vortex rings  placed randomly in the 
periodic box of size $D$ as a function of the rings' radius $R$. 
We find that if $R \ge D$ (in which case rings are ``folded'' into broken
arches by
the periodic boundary conditions)
most of the energy is concentrated at the largest length scales, 
whereas if $R < D$ the energy spectrum peaks at intermediate
scales, in analogy with the counterflow spectrum.

\section{Coherent structures}
\label{section:5}

\begin{figure}[h]
\begin{center}
\includegraphics[width=0.50\textwidth]{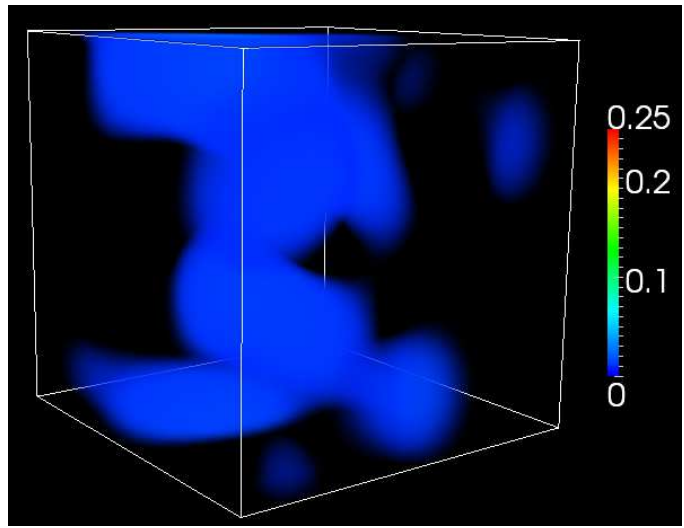}
\caption{
(Color online)
Smoothed vorticity $\bom$ sustained by a constant $\bv_n^{ext}$ (thermally-driven turbulence).
}
\label{fig:6}
\end{center}
\end{figure}

\begin{figure}[h]
\begin{center}
\includegraphics[width=0.50\textwidth]{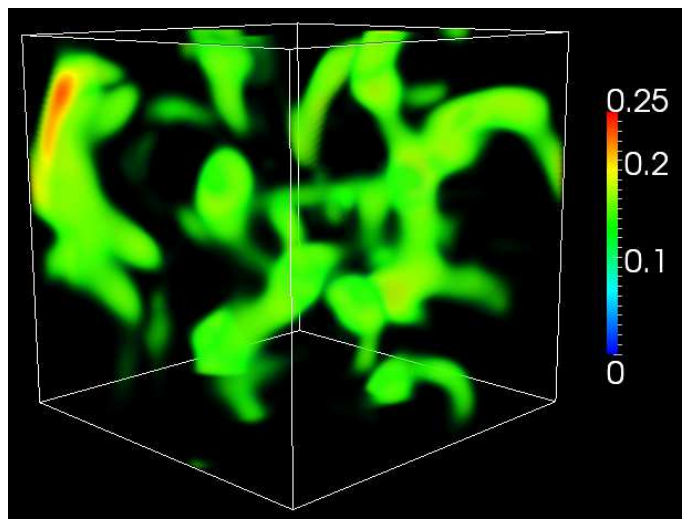}
\caption{
(Color online)
Smoothed vorticity $\bom$ sustained by a turbulent $\bv_n^{ext}$ 
(mechanically-driven turbulence). 
Notice the intense vortical regions compared to 
Fig.~\ref{fig:6} which is plotted on the same scale.
}
\label{fig:7}
\end{center}
\end{figure}

We also notice another difference between the two forms of turbulence.
If we convolve the vortex filaments
with a Gaussian kernel and define a smoothed vorticity
field $\bom_s$ (the details of the procedure are
described  in Ref.~\cite{Baggaley-structures}), it becomes
apparent - see Fig.~\ref{fig:6} -  that
the thermally-induced tangle (sustained by the uniform $\bv_n^{ext}$)
is essentially featureless, whereas the
mechanically-induced tangle (sustained by the turbulent $\bv_n^{ext}$)
contains ``vortical worms'', or  regions of concentrated vorticity
- see Fig.~\ref{fig:7}.
This result is consistent
with the observation of ``worms'' in two other related turbulent flows:
ordinary viscous turbulence \cite{Frisch}
and pure superfluid turbulence at $T=0$ without the normal fluid 
\cite{Baggaley-structures}; both flows
satisfy the Kolmogorov $k^{-5/3}$ scaling. 

It is known from previous work
that if intense regions of normal 
fluid vorticity are imposed, such as Gaussian
vortex tubes \cite{Samuels}, ABC flows \cite{Bauer} or 
worms \cite{Kivotides,Morris}, these structures will
induce (via the friction force) similar structures in the superfluid
vortex lines.
Our synthetic turbulent flow $\bv_n^{ext}$, although not
completely featureless on its own, contains only weak vortex structures,
much smaller \cite{Fung} than the vortical worms arising from direct
numerical simulations of the Navier-Stokes equation.
Therefore the observation of superfluid vortex bundles 
driven by the synthetic turbulent flow $\bv_n^{ext}$ of Eq.~(\ref{eq:KS})
must be an underestimate of the strength of these bundles. 
If we solved the Navier-Stokes equation for the normal fluid (rather
than imposing $\bv_n^{ext}$), the normal fluid's worms would probably
"imprint" vortex bundles in the superfluid, besides the bundles which 
arise naturally in the superfluid
as a consequence of Euler dynamics \cite{Baggaley-structures}.

A tentative explanation of the observation that the vortex configuration is
rather homogeneous for thermally-driven turbulence and inhomogeneous
for mechanically-driven turbulence is that
in the former (assuming, as we do, a uniform
normal flow) the growth rate of the Donnelly-Glaberson (DG) instability 
(which transforms normal fluid's energy into superfluid vortex length)
is the same everywhere, whereas in the latter it changes with time and space.

The DG mechanism is the following \cite{Tsubota-rot}.
If it is large enough, the component $V$ of the normal fluid velocity 
along a vortex lines can destabilise a (helical) Kelvin wave of given
wavenumber $k$. In this case,
the Kelvin wave grows with amplitude ${\cal A}(t) ={\cal A}(0) e^{\sigma t}$,
where $A(0)$ is the initial amplitude of the helix and
\begin{equation}\label{eq:DG_sigma}
\sigma(k)=\alpha (k V-\nu' k^2)
\end{equation} 
is the growth rate,
$\nu'=\kappa {\cal L}_1/(4 \pi) \approx \kappa$, and
${\cal L}_1=\ln{\left[1/(k a_0)\right]}$.
The growth of the Kelvin wave, however, may be 
interrupted by a vortex reconnection which ``breaks" the vortex line. 
It is known that vortex
reconnections play an essential
role in the turbulence \cite{Feynman,Schwarz,Nemirovskii-kinetics}. 
Therefore, it is prudent to assess the effect of reconnections on the 
DG instability. 

Consider mechanically-driven turbulence in a statisticaly steady state
at $T=1.9~\rm K$ ($\alpha=0.206$) driven by the rms normal fluid
velocity $V \approx 0.93~\rm cm/s$, with average vortex length
$\Lambda\approx 11.5~\rm cm$, vortex line density 
$L \approx 1.15 \times 10^4~\rm cm^{-2}$, and intervortex spacing 
$\ell \approx 9.3 \times 10^{-3}~\rm cm$, and thermally-driven
turbulence at the same temperature with $V=0.75~\rm cm/s$, $\Lambda \approx 11.88~\rm cm$,
$L \approx 1.19 \times 10^4~\rm cm^{-2}$, and 
$\ell \approx 9.2 \times 10^{-3}$. 
The average number $\zeta$ of vortex reconnections
per unit time is monitored during the numerical calculations; we obtain
$\zeta \approx 4370$ and  $7386~\rm s^{-1}$ for mechanically and thermally
driven turbulence, respectively (in reasonable agreement with the estimate
$\zeta \approx (2/3) \kappa L^{5/2} \approx 9500~\rm s^{-1}$, 
for a homogeneous isotropic tangle, of Barenghi \& Samuels \cite{Barenghi-Samuels}). 

The  mode which undergoes the most rapid DG instability has wavenumber
$k_{\rm max}=V/(2 \nu')$ and growth rate $\sigma_{\rm max}=\alpha V^2/(4 \nu')$,
corresponding to the
length scale $d_{\rm max}=2 \pi/k_{\rm max}$. In both mechanically-driven
and thermally-driven cases this
length scale  ($d_{\rm max}=0.015$ and
$0.017~\rm cm$ respectively) is larger than than the average distance between
vortices $\ell$, and so not suitable for our analysis. We therefore perform an analysis
for Kelvin waves with a wavelength and a wavenumber equal to $\ell$ and $k_\ell=2\pi/\ell$, respectively; we assume that such waves are the lowest frequency waves in our system. The growth rate of such a waves is $\sigma_{DG}=\sigma(k_\ell)$, where $\sigma(k)$ is defined in Eq.~(\ref{eq:DG_sigma}).

\begin{figure}[h]
\centering
\includegraphics[width=0.45\textwidth]{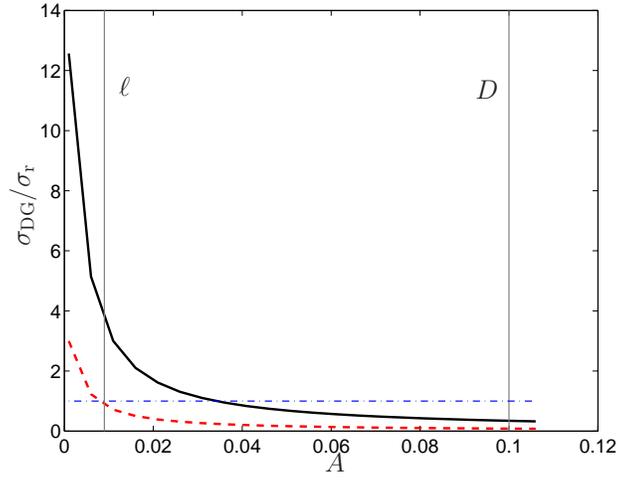}
\caption{(Color online) Plot of $\sigma_{DG}/\sigma_r$ (ratio of Donnelly-Glaberson and
vortex reconnection frequencies) vs wave amplitude $A$ (cm), for thermally (dashed, red line) and mechanically (solid line) driven turbulence; the (blue) dot-dashed line represents $\sigma_{DG}=\sigma_r$.
}
\label{fig:8}
\end{figure}

The reconnection rate $\zeta$ computed during the simulations is a statistical property of the vortex tangle as a whole. However we can compute a reconnection frequency for a  wavenumber and wave amplitude by scaling the total reconnection rate by the fraction of the total vortex length that a given wavelength takes. In such a manner we define
\begin{equation}
\sigma_r({\cal A})=\dfrac{\zeta}{\Lambda}\int_0^\ell \left [1+({\cal A}k_\ell)^2\cos^2(k_\ell x) \right ]^{-1/2}dx.
\end{equation} 
Fig.~\ref{fig:8} shows the ratio $\sigma_{DG} / \sigma_r$ vs wave amplitude
${\cal A}$ for the two simulations described above. We note a large contrast in the behaviour of the ratio of these two timescales when comparing the mechanically and thermally driven cases. For the latter one would estimate that the amplitude of perturbations along the vortices can grow to approximately the intervortex spacing before reconnections dominate the behaviour of the tangle. However, in the mechanically driven case $\sigma_{DG} \approx \sigma_r$ for ${\cal A} \approx D/3$ so that the large amplitude perturbations are able to grow, before reconnections randomise the tangle and introduce topological changes. Therefore the difference in the balance between these two competing timescales is likely to be partially responsible for the differences in the nature of the two turbulent systems.

\section{Vortex reconnections}
\label{section:6}

The existence of superfluid vortex bundles
\cite{Samuels,Bauer,Kivotides,Morris}, their dynamics \cite{Alamri}
and their particular significance at very low temperatures
\cite{L'vov2007,Kozik:2008} have been discussed in the 
literature, but so far there is no clear experimental evidence
for them.
It has been argued that the presence of bundles
of locally almost parallel vortices (which we have demonstrated in
the previous section for mechanically-induced turbulence)
leads to a suppression of vortex reconnections \cite{L'vov2007,Kozik:2008}.

\begin{figure}[h]
\begin{center}
\includegraphics[width=0.37\textwidth]{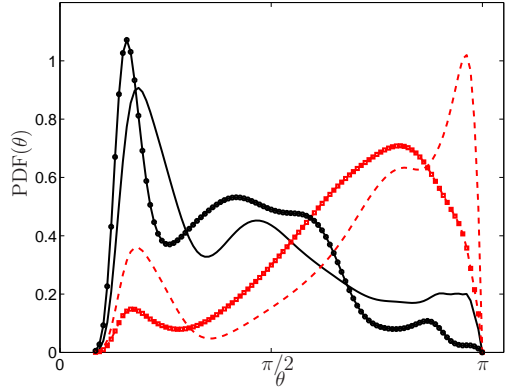}
\caption{(Color online) The probability density function (PDF) 
of the angle between reconnecting vortices,  $\theta$.
Thermally-driven turbulence: at $T=2.1~K$ (dashed red line) and 
at $T=1.9~K$ (solid red squares);
mechanically-driven turbulence: at $T=2.1~K$ (solid black line) and
at $T=1.9~K$ (solid black line with solid black circles). Note that
for thermally-driven turbulence the distribution peaks at large $\theta$,
whereas for mechanically-driven turbulence it peaks at small $\theta$.
}
\label{fig:9}
\end{center}
\end{figure}

In the vortex filament model, vortex reconnections are performed
algorithmically; the details are
described in Ref.~\cite{Baggaley-reconnections}. 
Within the approximation, intrinsic to the model, 
it is instructive to study the distribution of the angles $\theta$ between
reconnecting vortex lines at the level of discretization which we
use (which is necessarily much larger than $a_0$). 
The normalised distribution of values of $\theta$,
${\rm PDF}(\theta)$, is shown in Fig.~\ref{fig:9}: the solid black line
with black circles refers to mechanically-driven turbulence, and the solid
red line with red squares to thermally driven turbulence. 
It is apparent that in thermally-induced turbulence the majority of vortex
reconnections take place between vortex filaments 
which are nearly anti-parallel ($\theta \approx \pi$), whereas 
in mechanically-driven turbulence most reconnections are between vortices 
which are nearly parallel ($\theta<\pi/2$).  
Our results confirm that indeed the presence of organised bundles of 
vortices changes the typical geometry of reconnections.

To check the temperature dependence of the results we repeat our
calculations at higher temperature, $T=2.1~K$. At this temperature
the friction coefficients are larger
($\alpha=1.21$ and $\alpha'=-0.3883$), therefore a more intense vortex
tangle is generated at the same value of the drive; moreover, 
short Kelvin waves are damped out more quickly.
We check that at the higher temperature $T=2.1~K$ the same differences
between thermally-driven and mechanically-driven turbulence are
present, which we
have described in the previous sections for $T=1.9~\rm K$ 
in terms of
energy spectrum, curvature and coherence structures. Fig.~\ref{fig:9}
shows that, qualitatively, the distribution of reconnecting angles is also
temperature independent
(the black solid line, which refers to mechanically-driven turbulence peaks
at small $\theta$, the solid red line which refers to thermally-driven
turbulence peaks at large $\theta$).

\begin{figure}[h]
\centering
\begin{tabular}{cc}
\includegraphics[width=0.45\textwidth]{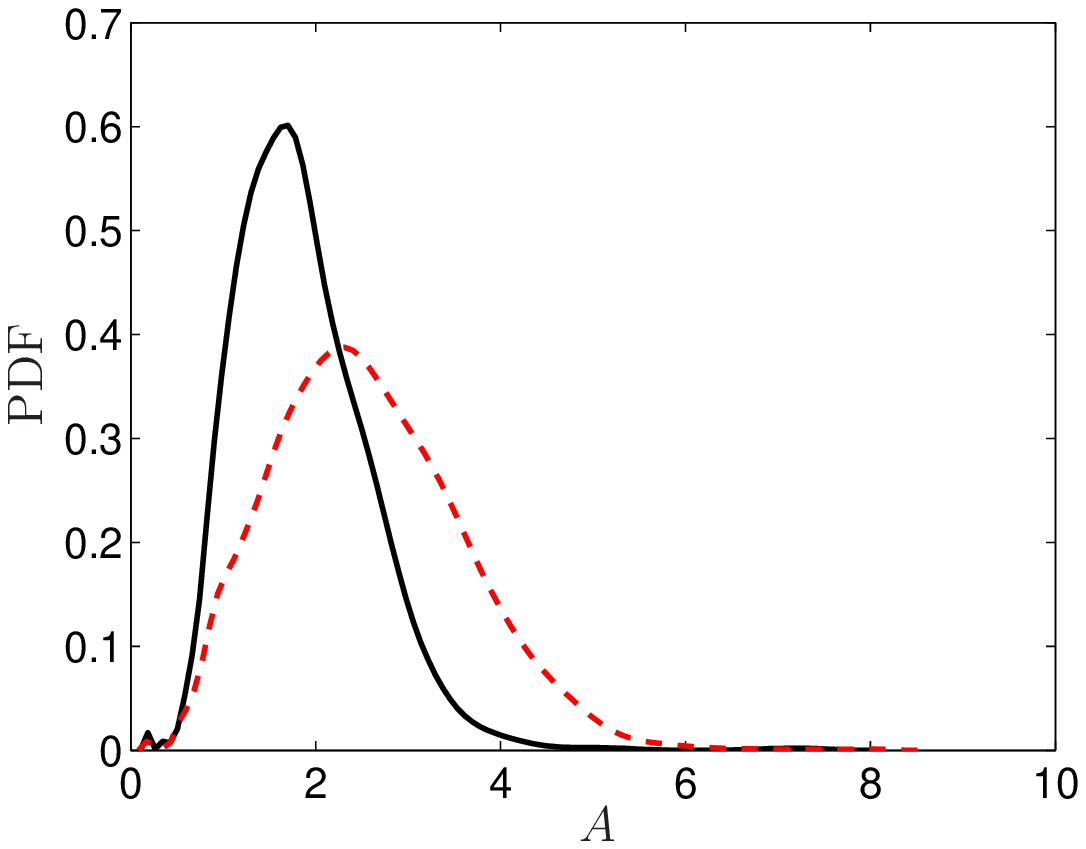}\\
\includegraphics[width=0.45\textwidth]{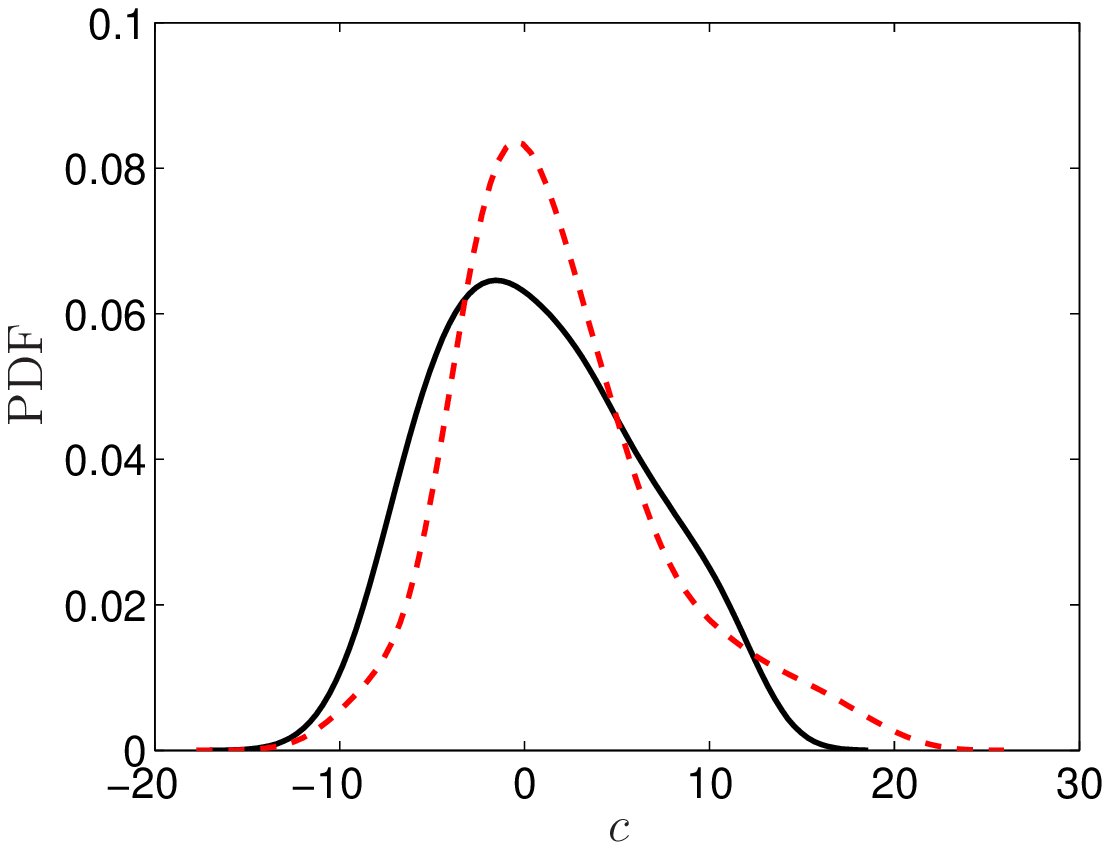}
\end{tabular}
\caption{(color online) Probability density functions (PDF) of 
the fitting parameters $A$ (top) and  $c$ (s$^{-1}$) (bottom)
of Eq.~\ref{eq:correction_factor}. 
Solid black line: mechanically-driven turbulence;
dashed red line: thermally-driven turbulence.
}
\label{fig:10}
\end{figure}

This result could be exploited to look for experimental evidence of
superfluid vortex bundles in the following way.
Using solid hydrogen tracer particles to visualise the vortex lines,
Paoletti, Fisher and Lathrop  \etal \cite{Paoletti} determined that the the minimum
distance $\delta(t)$ between vortex lines before and after a reconnection
scales as

\begin{equation}\label{eq:correction_factor}
\delta(t)=A(\kappa|t-t_0|)^{1/2}(1+c|t-t_0|),
\end{equation}

\noindent
where $t_0$ is the time at which the reconnection takes place,
with fitting coefficients $A \approx 1.2$ and $c \approx 0$.
We proceed in this way, monitoring vortex reconnections in our
numerical calculations.
Fig.~\ref{fig:10} shows the probability density functions of our 
fitting parameters $A$ and $c$ obtained for 1107 reconnections in thermally-driven 
turbulence (average values $\langle A\rangle=2.6$ and $\langle c\rangle=1.6$)
and 879 reconnections in mechanically-driven turbulence
(average values $\langle A\rangle=1.8$ and $\langle c\rangle=0.7\,{\rm s}^{-1}$).
Our fitting coefficients thus agree fairly well with
the experimental findings of Ref. \cite{Paoletti} 
and with the numerical results of
Tsubota and Adachi \cite{Tsubota-recon} ($A \approx 3$ and $c \approx 0\,{\rm s}^{-1}$).

\begin{figure}[h]
\centering
\begin{tabular}{ccc}
\includegraphics[width=0.45\textwidth]{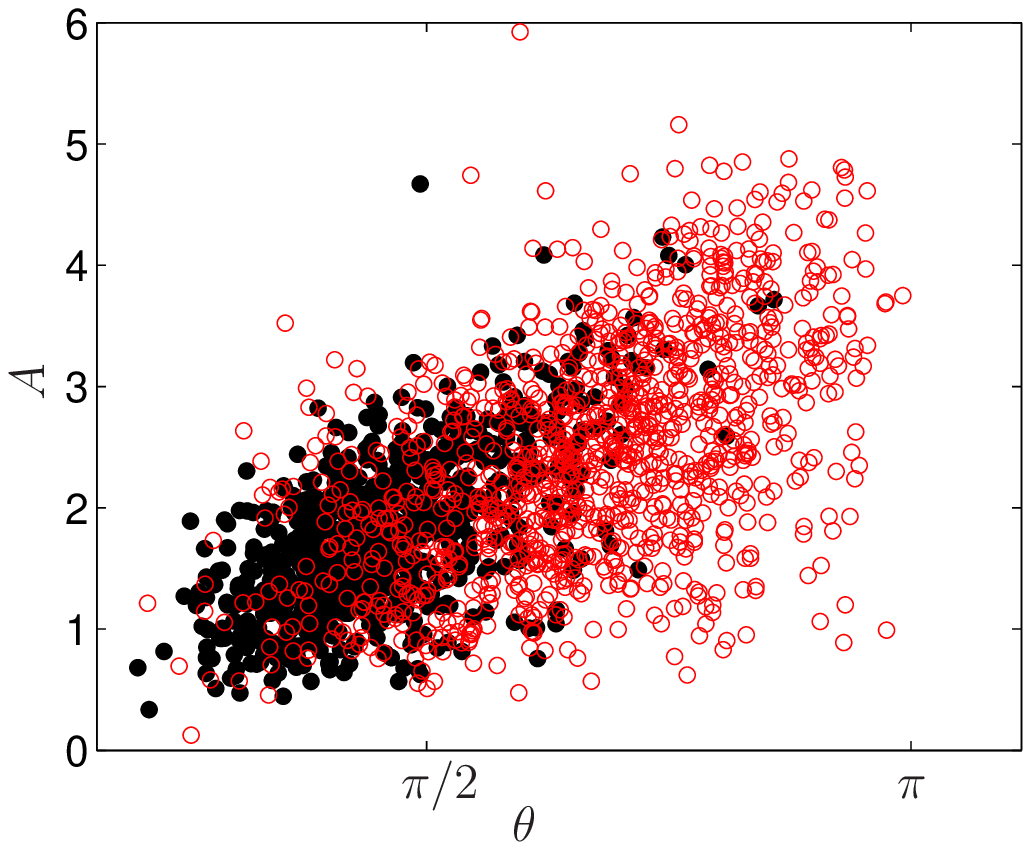}\\
\includegraphics[width=0.45\textwidth]{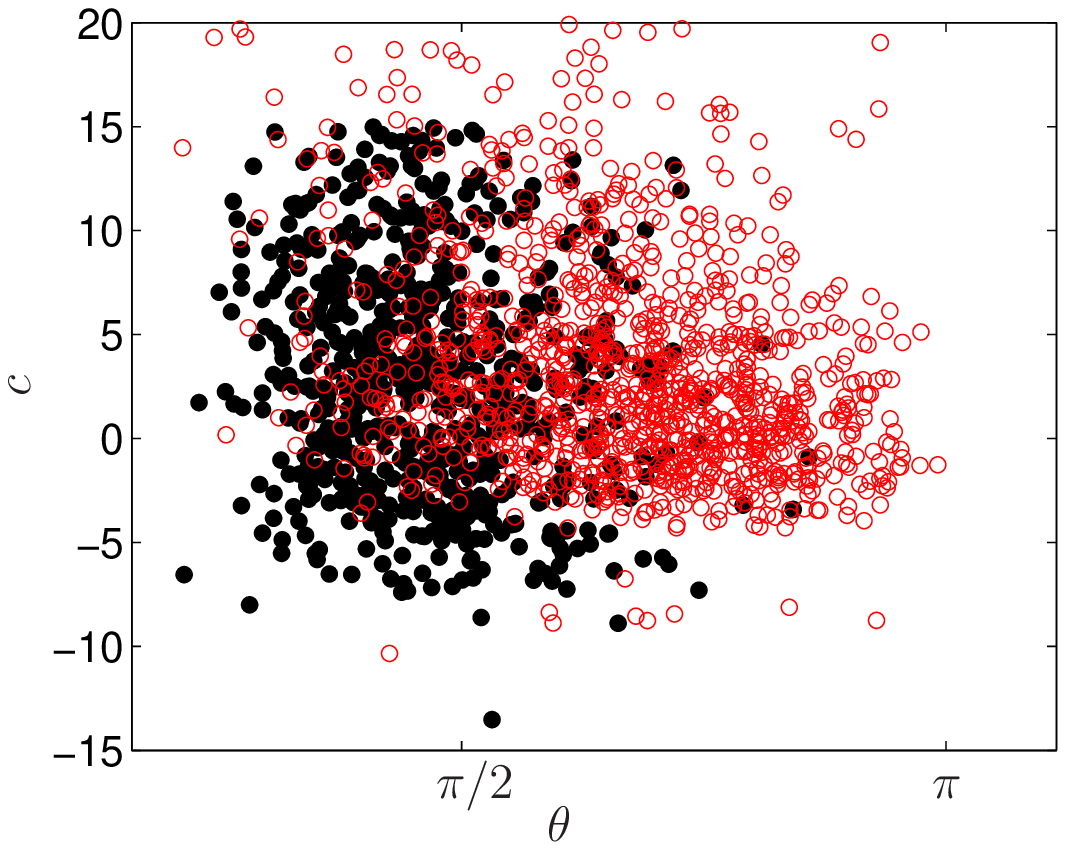}\\
\includegraphics[width=0.45\textwidth]{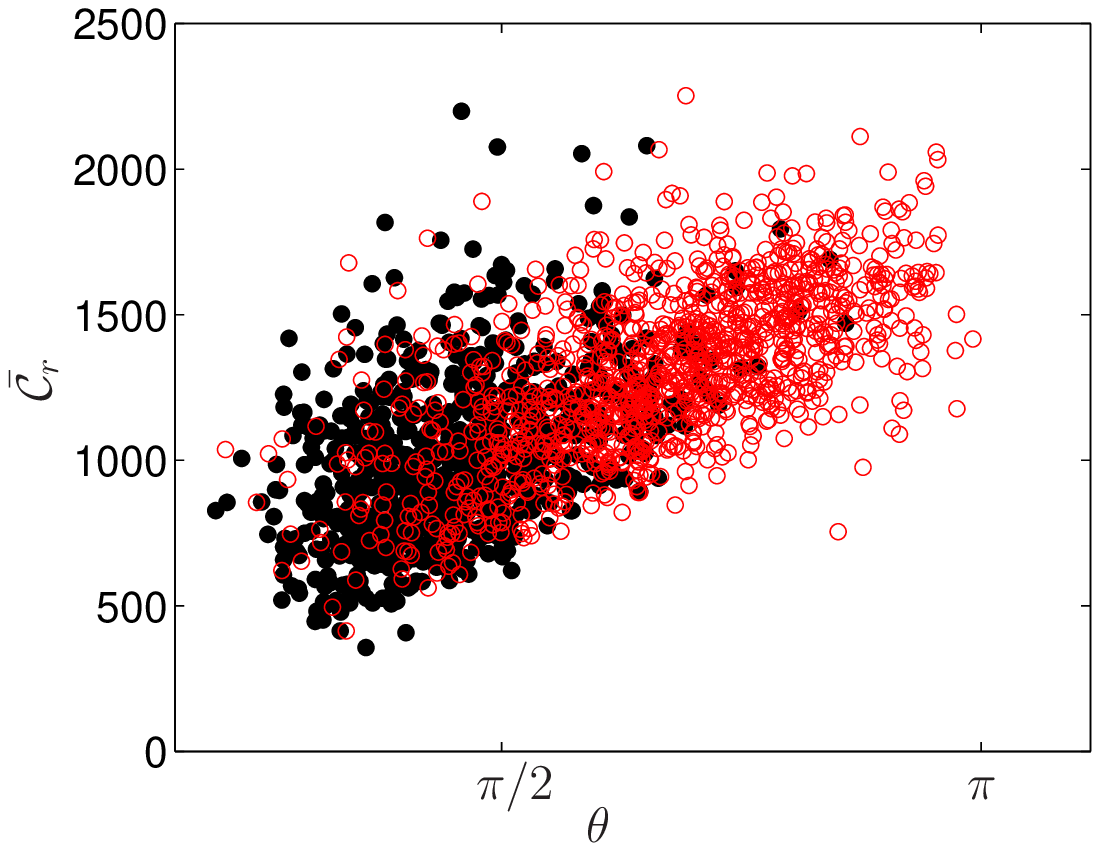}
\end{tabular}
\caption{(color online) Scatter plots of the 
fitting parameters $A$ (top) and $c$ (middle) 
of Eq.~\ref{eq:correction_factor}
vs the angle, $\theta$, between the 
reconnecting vortices. The bottom figure shows 
the mean curvature $\bar{\mathcal{C}_r}$ of
the reconnecting vortex segments vs the angle $\theta$.
Solid black points: mechanically-driven turbulence;
open red circles: thermally-driven turbulence.
}
\label{fig:11}
\end{figure}

Fig.~\ref{fig:10} shows that the distribution of values of $A$ is
different for thermally-driven and mechanically-driven turbulence.
The effect  must arise from the different distributions of curvature and
reconnecting angles $\theta$ for vortex bundles, which are present
only in mechanically-driven turbulence.
This is confirmed by Fig.~\ref{fig:11}, which displays scatter plots 
of the fitting parameters $A$ (top) and $c$ (middle). Fig.~\ref{fig:11}
also shows the angular dependence of the mean curvature 
$\bar{\mathcal{C}_r}$ (bottom) of the reconnecting vortex segments.

It is clear that the curvature of the filaments after a reconnection
is dependent on the angle of the reconnection.
From inspection of the local induction approximation \cite{Saffman} we would expect larger velocities (and thus $A$) with increased curvature, as we observe in the numerical simulations.
These result suggests a possible experimental strategy to 
establish the existence of vortex bundles based on the careful 
analysis of the reconnection fitting parameter $A$.

\section{Conclusions}
\label{section:7}

In conclusion, we have addressed for the first time the question of the energy
spectrum of thermally-induced counterflow turbulence, and found that it is 
unlike the spectrum of turbulence generated mechanically.
More in general, we have found that the two forms of quantum turbulence 
which can be generated in superfluid helium are quite different. 
Counterflow turbulence,
driven thermally by a constant normal fluid velocity, is uniform in
physical space and the energy spectrum is concentrated  at intermediate 
wavenumbers $k$.
On the contrary, quantum
turbulence driven mechanically by a turbulent normal fluid contains
regions of concentrated coherent vorticity and
vortex lines with larger radii of curvature; 
the energy is concentrated at the largest scales,
exhibiting the same $k^{-5/3}$ scaling of ordinary turbulence which
suggests the presence of an energy cascade. Our results prove that
counterflow turbulence, a form of disordered heat transfer unique to
liquid helium, lacks the
multitude of interacting length scales which is perhaps the main
property of ordinary turbulence. Vortex reconnections are affected
by the presence of bundles of almost parallel vortices, 
suggesting an experimental
technique to detect these bundles based on monitoring the vortex
separation after reconnections.

\begin{acknowledgments}
We thank the Leverhulme Trust and the EPSRC for financial support.
\end{acknowledgments}

\newpage

\end{document}